\documentclass[aps,prl,twocolumn,groupedaddress]{revtex4}  
\usepackage{graphicx,color}  
\usepackage{dcolumn}   
\usepackage{bm}        
\usepackage{amsfonts,amssymb,amsthm,amsmath}
\usepackage{graphicx}
\usepackage{float}
\usepackage{subfig}

\begin{document}



\title{On the edge energy of graphene}

\author{Tim R. Krumwiede\footnote{Department of Mathematics, Brigham Young University}}
\author{Tim P. Schulze\footnote{Department of Mathematics, University of Tennessee}}%
\affiliation{ }

\date{\today}

\begin{abstract} 
 Surface/edge energy is typically modeled as a continuous function of
  orientation, $\gamma({\bf n})$.  We put forward a simple geometric
  argument that suggests this picture is inadequate for crystals with
  a non-Bravais lattice structure. In the case of the idealized graphene/hexagonal lattice, our
  arguments indicate that the edge energy can be viewed as both
  discontinuous and multi-valued for a subset of orientations that are
  commensurate with the crystal structure.
\end{abstract}

\maketitle 

Surface/edge energy is normally modeled as a continuous function of
surface/edge orientation, $\gamma(\hat{{\bf n}})$ \cite{langer, BWBK}. This function is often
constructed so that it is consistent with underlying symmetry
constraints combined with experimental observations or data from
computations \cite{elder, MBW}.  In this paper, we put forward a simple geometric
argument that suggests this picture is incomplete for crystals with a
non-Bravais lattice structure.  Instead, we argue that when such a
crystal is cleaved by an arbitrarily positioned plane or line, the
resulting surface energy can be both multivalued and discontinuous
when viewed as a function of orientation.  This is due to the fact
that that some orientations give rise to a translation invariant
surface energy, while others do not.


These singularities occur for a discrete set of orientations that are
commensurate with the crystal structure. In the case of graphene, this
includes the so-called ``zigzag'' orientation, which is often found to
dominate the equilibrium shape of isolated graphene crystals.  While
we expect similar conclusions to apply to real materials and
first-principles calculations, including graphene films grown on
substrates, we examine these effects for an isolated hexagonal lattice
using nearest-neighbor, bond-counting arguments, neglecting
reconstructions and other off-lattice effects. The
  surface energy we discuss throughout most of the paper corresponds
  to the zero-temperature energy landscape of an ideal bulk-truncated
  surface/edge.

In a pairwise bond-counting model, an energy is defined for a given
lattice configuration by defining sets of bond orientations ${\rm
  V}=\{\{{\bf v}_{ij}\}_{j=1}^{J_i}\}_{i=1}^{N_p}$ and corresponding
bond energies $e_{ij}$ for each of the $N_p$ particles in the system
\cite{MMN}.  These sets are often restricted to neighboring pairs of
atoms, but, in principle, could include all combinations of atoms. For
our nearest-neighbor hexagonal/graphene model, the particles have one
of two distinct sets of bonds, $\{\{{\bf v}_{Aj}\}_{j=1}^{3},\{{\bf
  v}_{Bj}\}_{j=1}^{3}\}$.  For a crystal with a Bravais lattice
structure, the same set of bonds, $\{\mathbf{v}_j \}_{j=1}^J$, applies
to each particle in the crystal.  The surface/edge energy of
bond-counting models on Bravais lattices is given by
\begin{equation}
  \gamma(\hat{\mathbf{n}})=\frac12\sum_{j=1}^J e_j \frac{|\hat{\mathbf{n}}\cdot\mathbf{v}_j|}{|\det A|}
\end{equation}
where $\hat{\mathbf{n}}$ is the normal to the surface/edge
 and $A$ is a matrix
with the lattice primitive vectors as columns \cite{herr}.

The idealized graphene structure is one of the simplest examples of a
non-Bravais lattice.
Gan and Srolovitz were the first to address the issue of edge
energy for individual graphene flakes \cite{GS}. They use 
DFT calculations for a collection of graphene ribbons at seven
different orientations to interpolate an edge energy function, and consider unreconstructed graphene with both
non-terminated and hydrogen terminated bonds, as well as a model for
reconstructed graphene. Liu et al. \cite{LDY} revisit the problem and first 
consider an arbitrarily oriented graphene edge that can be decomposed into
a number of ``zigzag'' and ``armchair'' components, so that the edge energy
can be represented using two energies of these primary configurations
along with zigzag and armchair densities that can be computed from
simple geometric considerations:
$$
\gamma(\chi)=\frac{4}{\sqrt3}\epsilon_A\sin(\chi)+2\epsilon_Z\sin(\pi/6-\chi),
$$
where $\epsilon_A$ and $\epsilon_Z$ are the energies of an atom in an armchair or zigzag component respectively and $\chi$ is the edge angle. 
This assumption is equivalent to assuming edges of the graphene flake do not contain singly-bonded
carbon-atoms. This same assumption appears to have been tacitly made
in \cite{GS}, as a perfectly linear edge with the slope indicated in their 
Fig. 2b would have an additional singly-bonded atom at the kinks 
along the edge. 
For the graphene structure, we will see that
including such atoms leads to a discontinuous, multi-valued edge
energy.


\begin{figure}
\includegraphics[width=\linewidth]{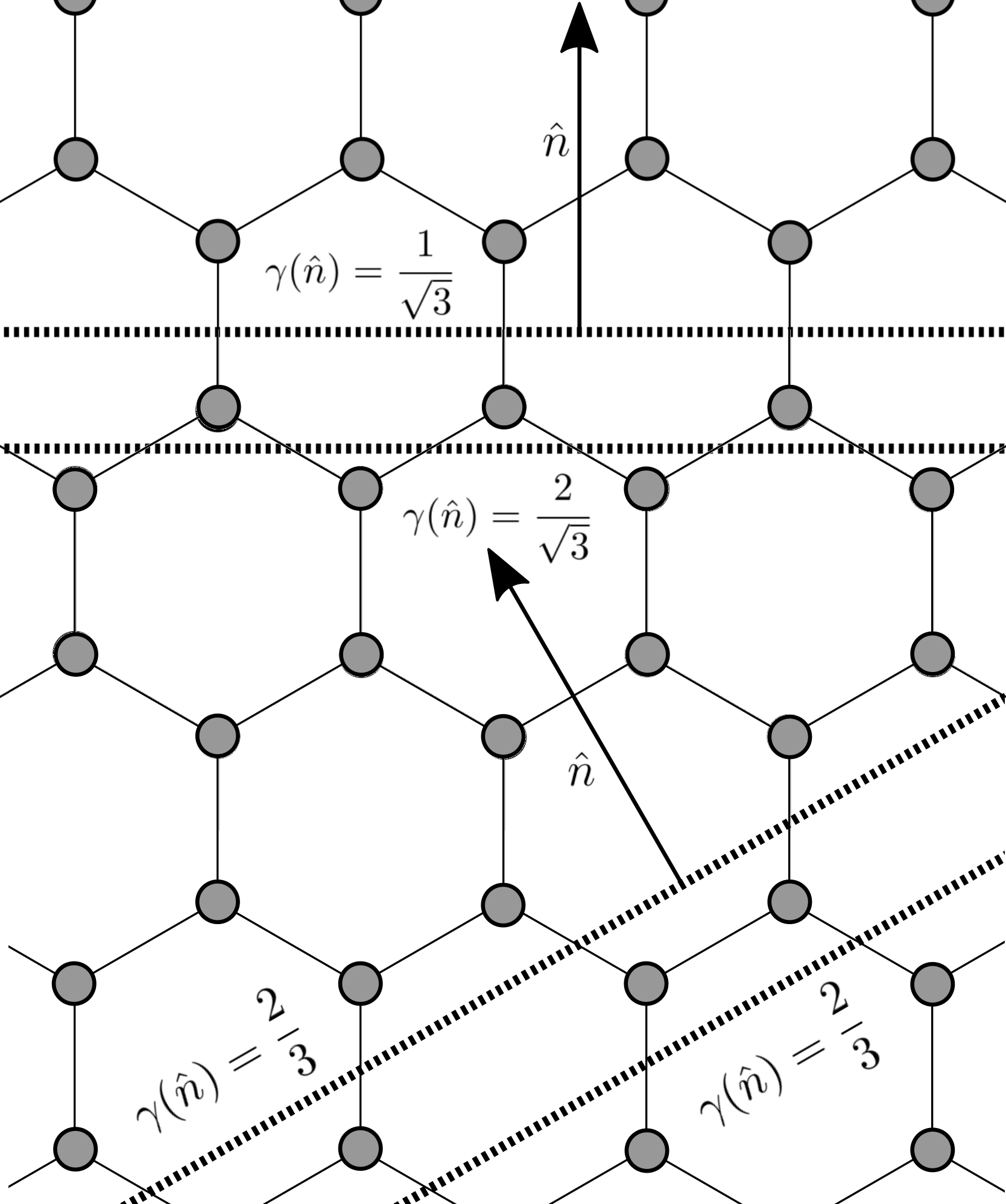} 
\hspace{ .2in}
\caption{\label{fig:epsart} A hexagonal/graphene lattice cut by two
  lines in the zigzag orientation (near the top of the figure) and two more
  in the armchair orientation (near the bottom of the figure). In the
  case of the zigzag orientations, the broken bond density can be
  altered by a parallel translation of the edge, while the broken
  bond density is translation invariant for the armchair
  orientation. }
\end{figure}

While it is not surprising that equilibrium shapes are dominated by
facets without these dangling atoms, it seems clear they would appear
in non equilibrium structures and could affect the dynamics of
relaxation and growth processes.  Indeed, Liu et al.  go on to
consider growth mechanisms involving singly-bonded carbon atoms
arriving at and diffusing along steps similar to what occurs in the
traditional Burton-Cabrera-Frank \cite{BCF} theory of step-flow on surfaces
, and singly-bonded atoms at graphene edges have been observed in experiments \cite{SK}.
In view of this, we examine a more complete picture of
surface/edge energy as a function of perfectly  planar/linear facets at
arbitrary orientations and positions.

The mechanism that is responsible for the discontinuities in the
surface energy is illustrated in Figure 1 using a nearest-neighbor
bonded crystal with the hexagonal/graphene structure. Most facets
behave like the armchair orientation shown at the bottom of Figure 1, where the
broken bond density, which represents edge energy in this simple
model, is translation invariant. This contrasts with a countable,
discrete set of orientations that behave like the zig-zag orientation
shown at the top of Figure 1, where the broken bond density alternates between
two values as the line cutting the crystal is translated in the normal
direction. 


\begin{figure}
\subfloat[]{\includegraphics[width=\linewidth]{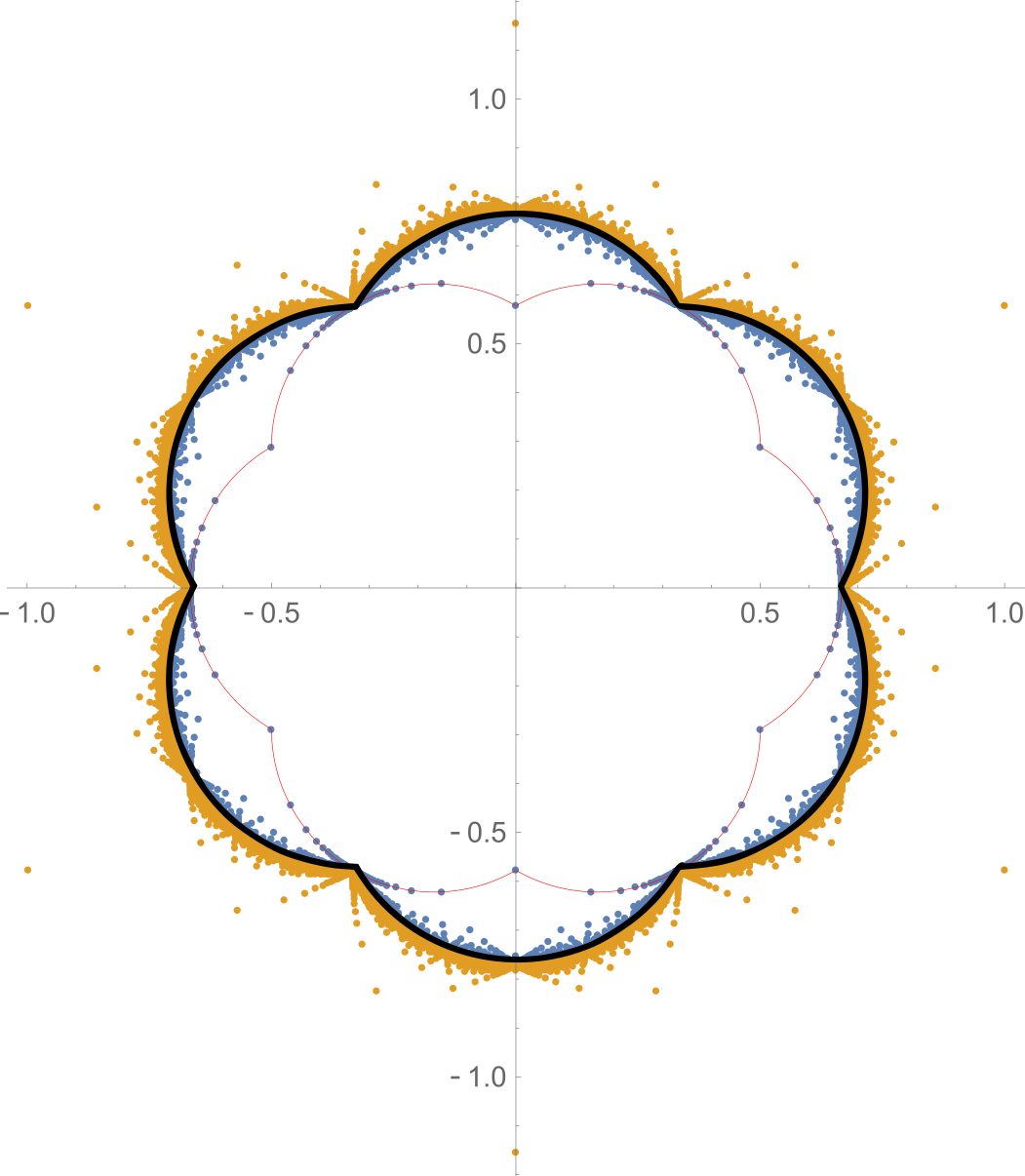}}

\subfloat[]{\includegraphics[width=\linewidth]{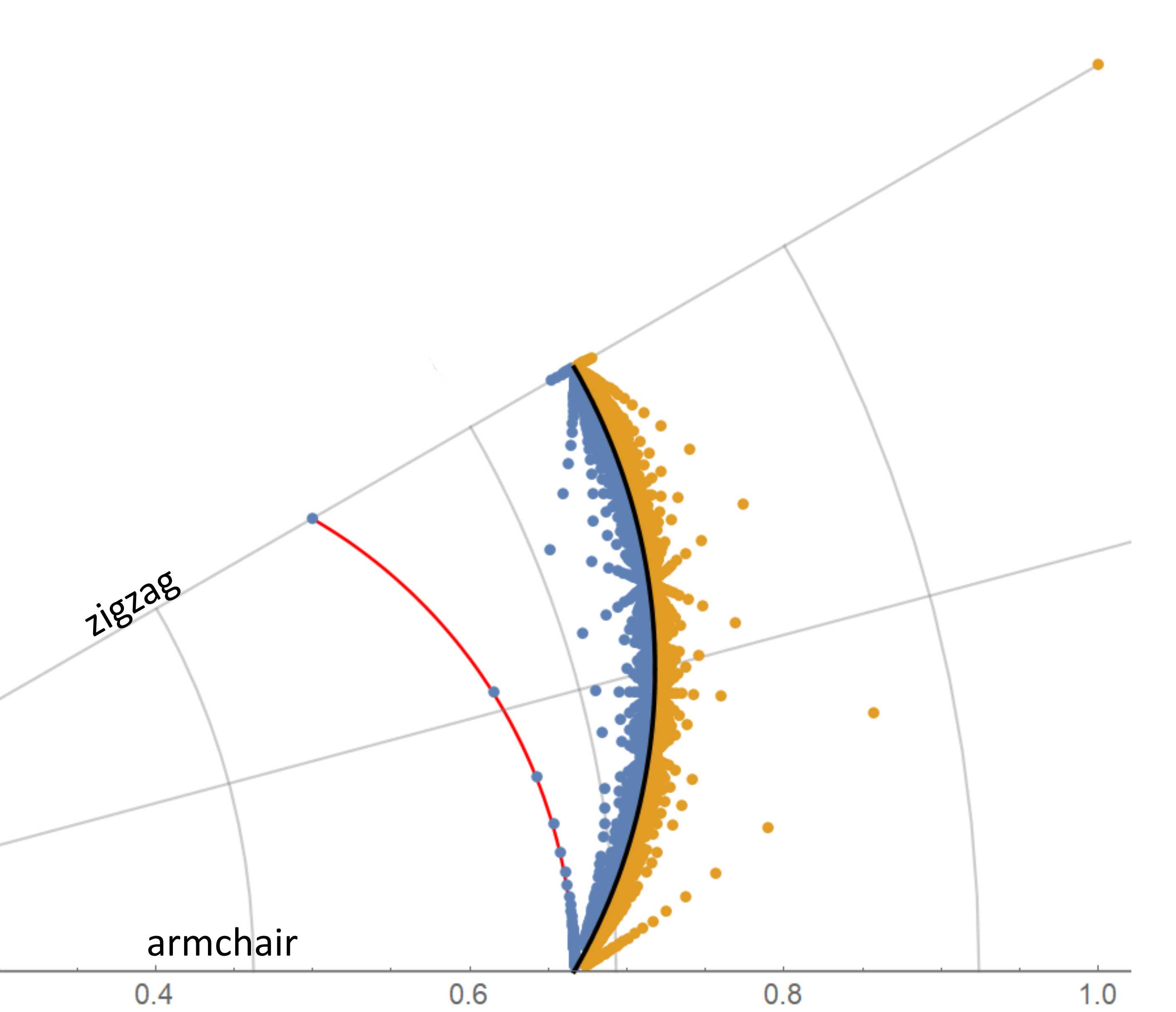}}

\caption{\label{fig:epsart} A polar plot of broken-bond density/edge
  energy for the hexagonal/graphene crystal as a function of edge
  orientation. The entire plot is shown in (a) while (b) contains only the wedge ranging between armchair and zigzag orientations. Most values lie on the black curve, while the
  discrete set of incongruent orientations gives rise to
  discontinuities with two edge energy values for each
  orientation: the minimum values are shown in blue and the maximum values are gold. The red curve interpolates between the zigzag and
  armchair orientation by neglecting dangling bonds. }
\end{figure}

In general, edge orientations fall into one of two categories: {\em
  commensurate} orientations result in a periodic pattern of broken
bonds, while {\em non-commensurate} orientations result in an
aperiodic pattern of broken bonds.
An edge with a commensurate orientation
can be translated so that it passes through multiple sites, while an
edge with an incommensurate orientation can pass through at most one
site. The commensurate edges give rise
to two subcases we refer to as {\em congruent} and {\em incongruent}.
While the incommensurate and congruent orientations have translation
invariant edge energies, the edge energy for the incongruent
orientations is multi-valued. 
An edge with a congruent orientation can be translated so that
it passes through either no sites or sites with both A-oriented and
B-oriented bonds, alternating between the two, while an edge with an
incongruent orientation can only pass through no sites or sites with
the same bond orientations.

These results are summarized in Figure 2. The black curve is the edge
energy that applies to the uncountably infinite number of
incommensurate and the countably infinite set of congruent edges.  This edge
energy is exactly $1/3$ what one would find for a nearest neighbor
model based on the related Bravais lattice with an additional lattice
point in the center of each hexagon.  This curve is discontinuous at
the incongruent orientations, where one finds two possible values of
the edge energy depending on the placement of the facet in the
normal direction. It can be shown that the average of these two values again
lies on the black curve. Finally, the red curve is the edge energy
derived in Liu et al. \cite{LDY} by assuming edges that consist of only armchair
and zigzag components. This curve is a lower bound on the edge
energy, and is formed by continuously interpolating between the lower of the
two possible values one can obtain with a zigzag orientation and the
single value for the armchair orientation.  
Note that if this simple,
interpolated edge energy function was used to evolve a non-equilibrium
shape, one would not expect any qualitative difference in the dynamics
compared to that for a material with a triangular lattice structure,
i.e. both edge energies are a six-petaled flower. 



\begin{figure}
\includegraphics[width=\linewidth]{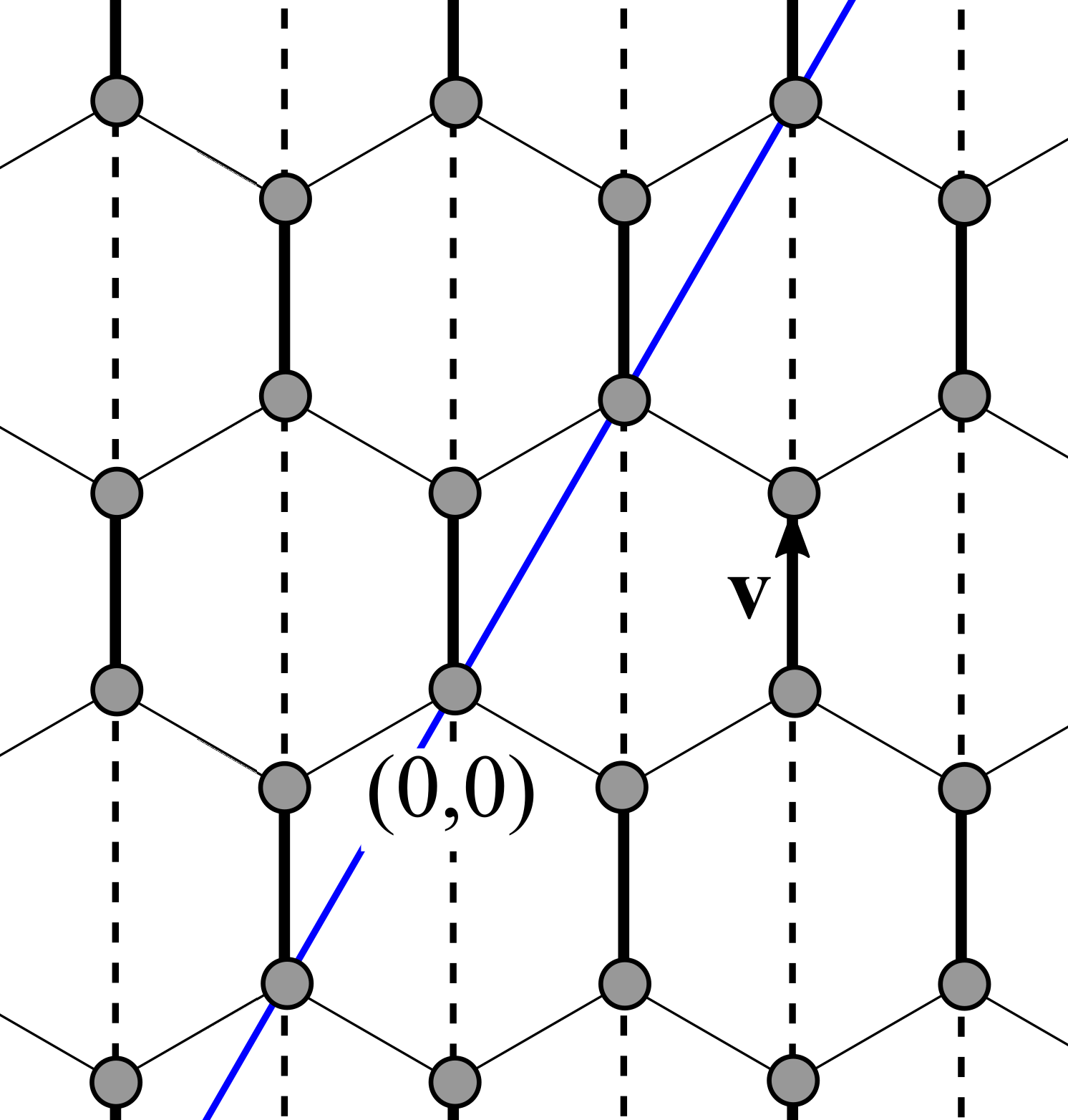}
\caption{\label{X} The hexagonal lattice with a single
  nearest-neighbor bond $\mathbf{v}$ and its corresponding set of
  bondlines. The blue reference line $\hat{y}=\sqrt{3}x$ is used to define
  the sequence $\delta_n$ referred to in the text.  }

\end{figure}

To get these results, we follow arguments that generalize those of
Mackenzie et. al. \cite{MMN}, i.e. we compute a contribution to the edge
energy for each bond orientation ${\bf v}$ and sum the result over all
bonds $V$.   To this end, consider the set of all {\em bond-lines} parallel
to $\mathbf{v}$ that pass through lattice sites.  Note that the bonds
only cover $\frac{1}{3}$ of each bond-line, with a repeating pattern
of one bond followed by two bond-less segments (see Figure 3). Thus, the
bonds and bond-line structure are periodic in the vertical direction
with period $3a$, where $a$ is the bond length.  We will make use of a
Cartesian coordinate system where the y-axis is aligned with the bond
and the origin is placed at the lower end of an arbitrary bond.
It is also convenient to introduce a reference line $\hat{y}=\sqrt{3} x$ and
measure distance in the direction of $\mathbf{v}$ relative to this
line, $\Delta=y-\hat{y}$.

Next, we consider an arbitrary edge $y=sx+b$ with slope $s$ and
intercept $b$. Between any two adjacent bond-lines, this line rises a
distance $r=\frac{\sqrt{3}}{2}as$ and intersects the $n^{th}$
bond-line, given by $x=n\frac{\sqrt{3}}{2}a$, at $y_n=rn+b$. Relative
to the reference line defined above, this produces the sequence
$\Delta_n=y_n-\hat{y}_n=(r-\frac{3a}{2})n+b$.  We will need to
consider y-values mapped to the interval $[0,3a]$ via congruence
modulo $3a$.  It will therefore be convenient to scale distance so
that $a=\frac{1}{3}$ and this congruence operation corresponds to
taking the fractional part of y-values. After 
 scaling, half of the bonds are  congruent to
the interval $[0,1/3]$ while the other half are congruent to the
interval $[1/2,5/6]$. Relative to the reference line,
the scaled bond locations will have
fractional parts in the interval $[0,1/3]$.
In
order to determine the intersections with bonds, it is sufficient to
consider the fractional part of the sequence
$\delta_n={\mbox{Frac}(\Delta_n)}$.

An edge orientation is commensurate with respect to bond $\mathbf{v}$
if $r\in \mathbb{Q}$ and {\em incommensurate} otherwise.  For a given
edge orientation, one can show that all of the bonds $\mathbf{v}\in V$
fall into the same category.  For incommensurate orientations, the
edge energy is defined as the mean number of bonds cut across the
entire edge.  A natural hypothesis for this mean is
$\gamma(\hat{\mathbf{n}})=\frac 13\Gamma(\hat{\mathbf{n}})$, where
$\Gamma$ is the edge energy for the related triangular lattice with
additional nodes in the center of each hexagon, as the graphene
lattice is formed by removing $\frac{2}{3}$ of these bonds.  The
triangular lattice is Bravais, so that $\Gamma$ can be computed from
(1).

To see that this is correct, we first consider the case $b=0$,
producing the sequence $\delta_n=\mbox{Frac}[(r-\frac{1}{2})n]$.
Since $r$ is irrational, so is $r-\frac{1}{2}$ and Weyl's
equidistribution theorem \cite{weyl} then indicates that the sequence
is uniformly distributed. This implies that one third of the bond-line
intersections correspond to broken bonds.  For $b \neq 0$,
$\mbox{Frac}[(r-\frac{1}{2})n+b] =
\mbox{Frac}[(r-\frac{3a}{2})n+\mbox{Frac}(b)]$,
from which we can see that broken bond density in the incommensurate
case is translation invariant, as the portion of the uniform
distribution of $\delta_n$ that is shifted out of the interval
$[0,1]$ on the right simply reemerges on the left.  The same result
holds for each $\mathbf v \in {\rm V}$ and therefore
\begin{equation}
  \gamma(\hat{\mathbf{n}})=\frac{1}{3}\Gamma(\hat{\mathbf{n}})=\frac{2}{3\sqrt{3}}\sum_{i=1}^{3}|\hat{\mathbf{n}}\cdot\mathbf{v}_i,
  |\end{equation} where we have expressed the result using twice the
  contribution from the three distinct A-bond orientations
  $\mathbf{v}_{A1}=(\sqrt{3}/2,1/2),
  \mathbf{v}_{A2}=(-\sqrt{3}/2,1/2), \mathbf{v}_{A3}=(0,1)$, as the
  B-bonds give rise to the same contributions.

The sequence $\delta_n$ is periodic whenever $r$ is rational,
repeating every $N$ bond-lines, where $N$ is the smallest even integer
such that $Nr \in \mathbb{Z}$. When this integer $N$ is divisible by
three, we refer to the orientation as {\em congruent}, as one can 
show that congruence applies to all bonds $\mathbf{v}\in V$ or none at all.
Congruent orientations have the same translation
invariant broken bond density as the incommensurate orientations. To
see this, note that the $N$ bond-line intersections are evenly spaced
over one period of length $p=rN$ and that the corresponding values of
$\delta_n$, though re-ordered, are uniformly spaced over the interval
$[0,1]$, with one third of these corresponding to a bond crossing.



The remaining commensurate cases have a repeating sequence
$\delta_n$ with $N \equiv1$ or $2\mod 3$.  In these cases,
which we refer to as {\em incongruent}, there is no way to have
exactly one third of the $\delta_n$ falling into the first third of
$(0,1]$.
Instead, the number
of intersections with bonds per period will round up or down to the
nearest integer that is divisible by 3.  If
$N \equiv 1\mod 3$,
 the lesser of the two
edge energies is given by
$$\gamma_{\mathbf {v}}^-(\hat{\mathbf{n}})=\frac{n-1}{3p}=\frac{p\Gamma_{\mathbf {v}}(\hat{\mathbf{n}})-1}{3p}=\frac13\Gamma_{\mathbf {v}}(\hat{\mathbf{n}})-\frac{1}{3p}, $$
and the greater of the two edge energies is
$$\gamma_{\mathbf {v}}^+(\hat{\mathbf{n}})=\frac13\Gamma_{\mathbf
  {v}}(\hat{\mathbf{n}})+\frac{2}{3p}.$$ Which of the two applies
depends on the intercept $b$ of the dividing line, and the edge energy
fails to be translation invariant for these orientations.  The
transition between the two values takes place whenever the edge
crosses a lattice site.  If $N\equiv 1\mod 3$, this occurs whenever
$\text{Frac}(b)=k/N$ or $k/N+1/(3N)$ with $k\in\mathbb{Z}$. And if
$N\equiv 2\mod 3$, when $\text{Frac}(b)=k/N$ or $k/N+2/(3N)$. We will
refer to the set of edges with $b$ between any two of these transition
values as a {\em band}. Note that all edges within a single band share the
same energy value. There are two possible values for any edge
orientation, with bands alternating between the two and one of the
bands being twice as wide as the other.

If $N\equiv 2\mod 3$, the two edge energy values are
$$\gamma_{\mathbf {v}}^-(\hat{\mathbf{n}})=\frac13\Gamma_{\mathbf {v}}(\hat{\mathbf{n}})-\frac{2}{3p},\;
\gamma_{\mathbf {v}}^+(\hat{\mathbf{n}})=\frac13\Gamma_{\mathbf 
  {v}}(\hat{\mathbf{n}})+\frac{1}{3p}.$$
The total edge energy for an edge within a thin band is
given by
\begin{eqnarray}
\gamma_1(\hat{\mathbf{n}})&=&\sum_{N\equiv 1\bmod 3}\gamma_{\mathbf{v}_i}^+(\hat{\mathbf{n}})+\sum_{N\equiv 2\bmod 3}\gamma_{\mathbf{v}_i}^-(\hat{\mathbf{n}}) \nonumber \\
&=&\frac13\Gamma(\hat{\mathbf{n}})+\frac{2}{3p}(m_1-m_2)
\end{eqnarray}
where $m_j$ is the number of bonds in $\{\mathbf{v}_i\}$ for which $N\equiv j\bmod 3$. Similarly, the edge energy for a edge within a thick band is
\begin{equation}
\gamma_2(\hat{\mathbf{n}})=\frac13\Gamma(\hat{\mathbf{n}})+\frac{1}{3p}(m_2-m_1).
\end{equation}
Note that
$$
\lim_{p \rightarrow \infty} \gamma_1 =\lim_{p \rightarrow \infty} \gamma_1 = \frac{1}{3}\Gamma,
$$ so that in the limit where the period of the bond intersections
becomes large, both of the values for the incongruent orientations
converge to the value for incommensurate/congruent orientations.

Let
$\gamma^-(\hat{\mathbf{n}})=\min\{\gamma_1(\hat{\mathbf{n}}),\gamma_2(\hat{\mathbf{n}})\}$
and
$\gamma^+(\hat{\mathbf{n}})=\max\{\gamma_1(\hat{\mathbf{n}}),\gamma_2(\hat{\mathbf{n}})\}$. These
two functions are then the minimum and maximum energy values for the
orientation $\hat{\mathbf{n}}$ shown in Figure 2.  In particular, the
zigzag orientations minimize $\gamma^-$ over 
all incongruent orientations.
 If we perform the classical Wulff construction using
$\gamma^-$ as the edge energy, we get a hexagon with zigzag
orientation edges, coinciding with the graphene equilibrium shape.


Further work examining the impact of these observations on the
nonequilibrium evolution of crystals would be of interest. In
particular, one would like to know if arbitrarily shaped crystals with
a hexagonal/graphene crystal structure exhibit behavior that is
qualitatively distinct from identically shaped crystals with a
triangular lattice structure. A possible mechanism for such
differences may be provided by the fact that a perturbation to the
energy minimizing orientation for the hexagonal/graphene crystal will
give rise to a jump in the edge energy, while a perturbation for the
energy minimizing orientation for a crystal with a Bravais lattice
structure will not.  Simulations using molecular dynamics, 
kinetic Monte Carlo and/or phase-field crystal would seem
well suited to exploring this possibility.

\begin{acknowledgments}
We wish to acknowledge support from NSF-DMS-1613729.

\end{acknowledgments}

\end{document}